\documentclass[onecolumn,floatfix,superscriptaddress,showpacs,showkeys,nofootinbib,preprint]{revtex4}

\usepackage{epsfig}
\usepackage{amssymb,latexsym,amsmath}
\newcommand{\eq}[1]{\begin{align} #1 \end{align}}

\begin{document}

\title{
Bose-Einstein Condensation in the Relativistic Pion Gas:\\
Thermodynamic Limit and Finite Size Effects }

 \author{V.V. Begun}
 \affiliation{Bogolyubov Institute for Theoretical Physics, Kiev, Ukraine}

 \author{M.I. Gorenstein}
 \affiliation{Bogolyubov Institute for Theoretical Physics, Kiev, Ukraine}
 \affiliation{Frankfurt Institute for Advanced Studies, Frankfurt,Germany}

\begin{abstract}
We consider the Bose-Einstein condensation (BEC) in a relativistic
pion gas. The thermodynamic limit when the system volume $V$ goes
to infinity as well as the role of finite size effects are
studied.
%We calculate the particle number
%fluctuations at the BEC and estimate the finite size effects.
At $V\rightarrow \infty$ the scaled variance for particle number
fluctuations, $\omega=\langle \Delta N^2\rangle/\langle N\rangle$,
%at the BEC is restricted by a finite system volume $V$. In the
%thermodynamic limit $V\rightarrow\infty$, the scaled variance
%$\omega$
converges to finite values in the normal phase above the
BEC temperature, $T>T_C$.
%In a region of the BEC,
It diverges as $\omega \propto V^{1/3}$ at the BEC line $T=T_C$,
and $\omega \propto V$ at $T<T_C$ in a phase with the BE
condensate.  Possible experimental signals of the pion BEC in
finite systems created in high energy proton-proton collisions are
discussed.

\end{abstract}


\maketitle

\section{Introduction}
Pions are spin-zero bosons. They are the lightest hadrons
copiously produced in high energy collisions. There were several
suggestions to search for the Bose-Einstein condensation (BEC) of
$\pi$-mesons (see, e.g., Ref.~\cite{pion-BC}). However, no clear
experimental signals were found up to now. Most of previous
proposals of the pion BEC signals were based on an increase of the
pion momentum spectra in the low (transverse) momentum region.
These signals appear to be rather weak and they are contaminated
by resonance decays to pions. In our recent paper \cite{bec2} it
was suggested that the pion number fluctuations strongly increase
and may give a prominent signal of approaching the BEC. This can
be achieved by selecting special samples of collision events with
high pion multiplicities.

In the present paper we study the  dependence of different
physical quantities on the system volume $V$ and, in particular,
their behavior at $V\rightarrow \infty$. As any other phase
transition, the BEC phase transition has a  mathematical meaning
in the thermodynamic limit (TL) $V\rightarrow\infty$. To define
rigorously this limit one needs to start with a finite volume
system. Besides, the finite size effects are important for an
experimental search of the pion BEC fluctuation effects proposed
in Ref.~\cite{bec2}. The size of the pion number fluctuations in
the region of the BEC is restricted by a finite system volume $V$.
To be definite and taking in mind the physical applications, we
consider the ideal pion gas. However, the obtained results are
more general  and can be also applied to other Bose gases.

The paper is organized as follows. In Section~II we consider the
BEC in the TL. Here we emphasize some specific effects of the BEC
in relativistic gases. Section~III presents a systematic study of
the finite size effects for average quantities and for particle
number fluctuations. In Section~IV we discuss the finite size
restrictions on the proposed fluctuation signals of the BEC in
high energy collisions with large pion multiplicities. A summary,
presented in Section~V, closes the paper.

\section{BEC in Thermodynamic Limit}\label{s-Pion-BEC}
\subsection{Phase Diagram}
\begin{figure}[h!]
\begin{center}
 \epsfig{file=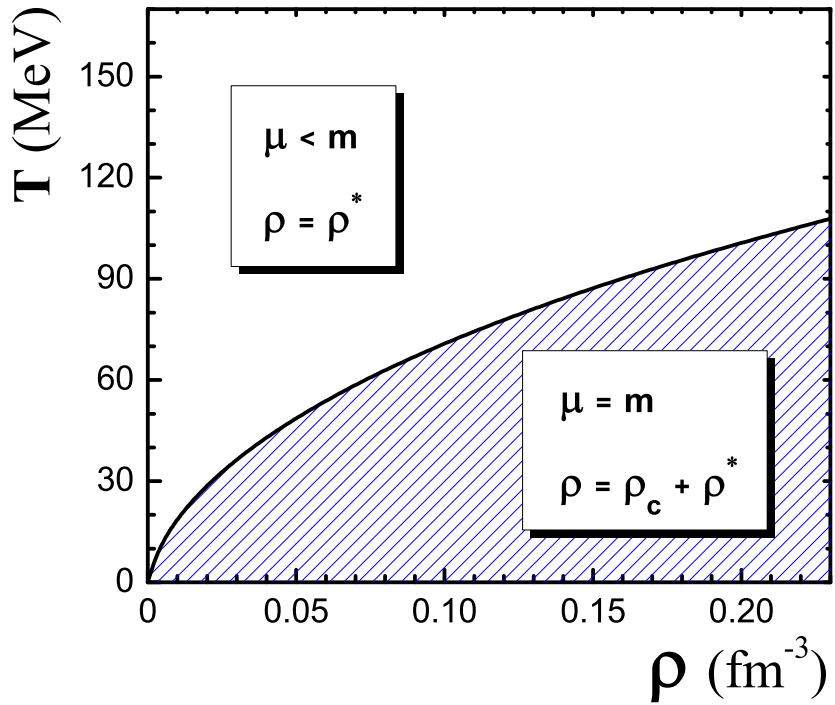,width=0.5\textheight}
 \caption{The phase diagram of the relativistic ideal pion gas with
 zero electric charge density $\rho_Q\equiv \rho_+-\rho_-=0$.
 The solid line shows the relation $T=T_C$ (\ref{T_BC}) of the BEC.
 Above this line, $T>T_C$ and there is a normal phase described by Eq.~(\ref{rho-pi}).
 Under this line, $T<T_C$ and there is a phase
 with the BE condensate described by Eq.~(\ref{rho-cond}). }
 \label{phase-diagr}
 \end{center}
\end{figure}

We consider  the relativistic ideal gas of pions. The occupation
numbers, $n_{\bf{p},j}$~, of single quantum states, labelled by
3-momenta $\bf{p}$, are equal to $n_{{\bf
p},j}=0,1,\ldots,\infty$~, where index $j$ enumerates 3 isospin
pion states, $\pi^+,\pi^-$, and $\pi^0$. The grand canonical
ensemble (GCE) average values, fluctuations, and correlations are
the following \cite{Landau},
 \eq{\label{np-pions} & \langle n_{{\bf
p},\,j}\rangle
 ~=~\frac{1}{\exp\left[\left(\sqrt{{\bf
 p^2}+m^2}~-~\mu_j\right) /T\right]~-~1}~,
 \\
 &\langle\left(\Delta n_{\bf{p},j}\right)^2\rangle
 \equiv  \langle \left( n_{{\bf p},j}-\langle n_{{\bf
p},j}\rangle\right)^2\rangle
 = \langle n_{\bf{p},j}\rangle \left(1
+ \langle n_{\bf{p},j} \rangle\right)\equiv
\upsilon^{2}_{\bf{p},j}~,
%\label{np-fluc}\\
~~~ \langle \Delta n_{\bf{p},j} \Delta n_{\bf{k},i} \rangle
 =\upsilon_{\bf{p},j}^2\delta_{\bf{p}\bf{k}}\delta_{ji}~,
  \label{np-corr}
 }
where the relativistic energy of one-particle states is taken as
$\epsilon_{{\bf p}}=({\bf p^2}+m^2)^{1/2}$ with $m\cong~140$~MeV
being the pion mass (we neglect a small difference between the
masses of charged and neutral pions), $T$ is the system
temperature, and chemical potentials are $\mu_+=\mu+\mu_Q$,
$\mu_-=\mu-\mu_Q$, and $\mu_0=\mu$, for $\pi^+$, $\pi^-$, and
$\pi^0$, respectively. In Eq.~(\ref{np-pions}) there are two
chemical potentials: $\mu_Q$ regulates an electric charge, and
$\mu$ a total number of pions.  In this paper we follow our
proposal of Ref.~\cite{bec2} and discuss a pion system with
$\mu_Q\!\!=\!0$ \footnote{The BEC in the relativistic gas of
`positive' and `negative' particles at $\mu_Q\rightarrow m$ has
been discussed in Refs.~\cite{Haber,Kapusta,Sal,Steph,bec1}}. This
corresponds to zero electric charge $Q$ which is defined by a
difference between the number of $\pi^+$ and $\pi^-$ mesons,
$Q=N_+-N_-=0$. The total pion number density is equal to:
 \eq{\label{rho-pi}
 \rho(T,\mu)
 &\;\equiv\;
 %\frac{\langle N_0+N_++N_-\rangle}{V}
 \rho_++\rho_+-\rho_0~=~\frac{\sum_{{\bf p},j}
 \langle n_{{\bf p},\,j}
 \rangle}{V}~\cong ~
  \frac{3}{2\pi^2}\;
        \int_0^{\infty}\frac{p^2 dp}{\exp \left[\left(\sqrt{p^{2}+m^{2}}
        \;-\;\mu\right)
              / T\right] ~-~ 1}~\nonumber
  \\
  & \equiv~\rho^*(T,\mu)
 \;=\; \frac{3\, T\,m^2}{2\,\pi^2}~
 \sum_{n=1}^{\infty}\frac{1}{n}\;K_2\left(n\,m/T\right)\,\exp(n\,\mu/T)~,
  }
where $\rho_j=\langle N_j\rangle/V$, with $j=+,-,0$, are the pion
number densities, $V$ is the system volume, and $K_2$ is the
modified Hankel function.  In the TL, i.e. for
$V\rightarrow\infty$, the sum over momentum states is transformed
into the momentum integral, $\sum_{\bf{p}}\ldots
=(V/2\pi^2)\int_0^{\infty}\ldots p^2dp$~. The particle number
density $\rho$ depends on $T$ and $\mu$, and volume $V$ does not
enter in Eq.~(\ref{rho-pi}). This is only valid at $\mu<m$. The
number of particles at zero momentum level is then finite, and its
contribution to particle number density goes to zero in the TL.
The inequality $\mu \le m$ is a general restriction in the
relativistic Bose gas, and $\mu=m$ corresponds to the BEC. The
Eq.~(\ref{rho-pi}) gives the following relation between the BEC
temperature $T_C$ and total pion number density $\rho$
\cite{bec2}:
 \eq{\label{T_BC}
%  \rho(T=T_C,\mu=m)~=~\rho^*(T=T_C,\mu=m)~
\rho~=~ \frac{3\,T_{C}\,m^2}{2\pi^2}
 \sum_{n=1}^{\infty}\frac{1}{n}\,
 K_2\left(n\,m/T_{C}\right)\,\exp(n\,m/T_{C})~.
  }
A phase diagram of the ideal pion gas in the $\rho-T$ plane is
presented in Fig.~\ref{phase-diagr}.

\noindent
 The line of the BEC phase transition is defined by
Eq.~(\ref{T_BC}), and it is shown by the solid line in
Fig.~\ref{phase-diagr}. In the non-relativistic limit,
$T_{C}/m\!\ll\!1$, using $K_2(x)\cong \sqrt{\pi/(2x)}\exp(-x)$ at
$x\gg 1$, one finds from Eq.~(\ref{T_BC}),
%\cite{Landau},
 \eq{\label{TC-nonrel}
%\rho(T=T_C,\mu=m)~\cong~3~\zeta(3/2)~\left(\frac{m~T_C}{2\pi}\right)^{3/2}
T_C~\cong~2\pi~[3\zeta(3/2)]^{-2/3}~m^{-1}~\rho^{2/3} ~\cong
~1.592~m^{-1}~\rho^{2/3},
 }
whereas in the ultra-relativistic limit\footnote{The BE condensate
formed in the ultra-relativistic regime has been considered in
Refs.~\cite{Madsen,cosm} as a {\it dark matter} candidate in
cosmological models.}, $T_{C}/m\!\gg\!1$, one uses $K_2(x)\cong
2/x^2$ at $x\ll 1$, and Eq.~(\ref{T_BC}) gives \cite{bec2},
 \eq{\label{TC-rel}
%\rho(T=T_C,\mu=m)~\cong~\frac{3\zeta(3)}{\pi^2}~T_{C}^3~.
T_C~\cong~[3\zeta(3)/\pi^2]^{-1/3}~\rho^{1/3}~\cong~1.399~\rho^{1/3}~.
}
In Eqs.~(\ref{TC-nonrel}-\ref{TC-rel}),
$\zeta(k)=\sum_{k=1}^{\infty}n^{-k}$ is the Riemann zeta function
\cite{Abr}, $\zeta(3/2)\cong 2.612$ and $\zeta(3)\cong 1.202$. The
equation (\ref{TC-nonrel}) corresponds to the well known
non-relativistic result (see, e.g., Ref.~\cite{Landau}) with pion
mass $m$ and `degeneracy factor' $3$.

The particle number density is inversely proportional to the
proper particle volume, $\rho \propto r^{-3}$. Then it follows
from Eq.~(\ref{TC-nonrel}) for a ratio of the BEC temperature in
the atomic gases, $T_C(A)$, to that in the pion gas, $T_C(\pi)$,
in non-relativistic approximation,
\eq{\label{T-pi}
\frac{T_C(\pi)}{T_C(A)}~\cong~
\frac{m_A}{m}~\left(\frac{r_A}{r_{\pi}}\right)^2~\cong~\frac{m_A}{m}~10^{10}~,
}
where $r_A\cong 10^{-8}$~cm and $r_{\pi}\cong 10^{-13}$~cm are
typical radiuses of atom and pion, respectively, and  $m_A$ is the
mass of an atom. The Eq.~(\ref{T-pi}) shows that $T_C(\pi)\gg
T_C(A)$, and this happens due to $r_{\pi}\ll r_A$.

 The equation (\ref{rho-pi}) gives the total pion number density
at $V\rightarrow\infty$ in the normal phase $T>T_C$ without BE
condensate. At $T<T_{C}$ the total pion number density becomes a
sum of two terms,
 \eq{\label{rho-cond}
 \rho ~=~\rho_{C}~+~\rho^*(T,\mu=m)~.
 }
The second term in the r.h.s. of Eq.~(\ref{rho-cond}) is given by
Eq.~(\ref{rho-pi}). The BE condensate $\rho_{C}$ defined by
Eq.~(\ref{rho-cond}) corresponds to a macroscopic (proportional to
$V$) number of particles at the lowest quantum level $p=0$.

To obtain the asymptotic expansion of $\rho^*(T,\mu)$ given by
Eq.~(\ref{rho-pi}) at $\mu\rightarrow m-0$ we use the identity
$[\exp(y)-1]^{-1}= [$cth$(y/2)-1]/2$ and variable substitution,
$p=\sqrt{2}m^{1/2}(m-\mu)^{1/2}x$, similar to those in a
non-relativistic gas \cite{Tolmachjev}. Then one finds,
\eq{\label{A1}
&\rho^*(T,m)~-~\rho^*(T,\mu)~=~\frac{3m^{3/2}}{\sqrt{2}\pi^2}(m-\mu)^{3/2}~
\int_0^{\infty}x^2dx\Big[\textrm{cth}\frac{\sqrt{2m(m-\mu)x^2+m^2}~-~m}{2T}~
\nonumber
\\&
-~
\textrm{cth}\frac{\sqrt{2m(m-\mu)x^2+m^2}~-~\mu}{2T}\Big]~\cong~
\frac{3m^{3/2}}{\sqrt{2}\pi^2}(m-\mu)^{3/2}~
\int_0^{\infty}x^2dx\Big[\textrm{cth}\frac{(m-\mu)x^2}{2T}~%\nonumber
\\&
- \textrm{cth}\frac{(m-\mu)(x^2+1)}{2T}\Big] \cong
\frac{6Tm^{3/2}}{\sqrt{2}\pi^2}(m-\mu)^{1/2}
\int_0^{\infty}x^2dx\Big[\frac{1}{x^2}-\frac{1}{x^2+1}\Big]
%\nonumber
%\\
=\frac{3Tm^{3/2}}{\sqrt{2}\pi}~(m-\mu)^{1/2}~.\nonumber
}
At constant density, $\rho^*(T,\mu)=\rho^*(T=T_C,\mu=m)$, one
finds in the TL at $T\rightarrow T_C+0$,
\eq{\label{A2}
&\rho^*(T=T_C,\mu=m)~=~\rho^*(T,\mu)~\cong~\rho^*(T,\mu=m)-\frac{3Tm^{3/2}}{\sqrt{2}\pi}(m-\mu)^{1/2}~
\nonumber \\
&\cong~\rho^*(T=T_C,\mu=m)~
+~\frac{d~\rho(T,\mu=m)}{dT}\Big|_{T=T_C}\cdot
(T-T_C)~-~\frac{3T_C~m^{3/2}}{\sqrt{2}\pi}~(m-\mu)^{1/2}~.
}
Using Eq.~(\ref{A2}) one finds the function $\mu(T)$ at
$T\rightarrow T_C+0$,
\eq{\label{A3}
m~-~\mu(T)~\cong
\frac{2\pi^2}{9T_C^2~m^3}\left[\frac{d~\rho(T,\mu=m)}{dT}\Big|_{T=T_C}\right]^2\cdot
(T-T_C)^2~,
}
In the non-relativistic limit $m/T \gg 1$ one finds,
$\rho(T,\mu=m)\cong 3 \zeta(3/2) \left[mT/(2\pi)\right]^{3/2}$,
similar to Eq.~(\ref{TC-nonrel}). Using Eq.~(\ref{A3}) one then
obtains \cite{Tolmachjev},
\eq{\label{A4}
\frac{m~-~\mu(T)}{T_C}~\cong~\frac{9\zeta^2(3/2)}{16\pi}~\cdot~
\left(\frac{T-T_C}{T_C}\right)^2~.
}
In the ultra-relativistic limit, $T/m \gg 1$, one finds,
$\rho(T,\mu=m) \cong 3\zeta(3)\;T^3/\pi^2$, similar to
Eq.~(\ref{TC-rel}). This gives,
\eq{\label{A5}
\frac{m~-~\mu(T)}{T_C}~\cong~\frac{18~\zeta^2(3)}{\pi^2}~\left(\frac{T_C}{m}\right)^3~\cdot~
\left(\frac{T-T_C}{T_C}\right)^2~.
}

Thus, $\mu(T)\rightarrow m$ and $d\mu/dT\rightarrow 0$ at
$T\rightarrow T_C+0$, both $\mu(T)$ and $d\mu/dT$ are continuous
functions at $T=T_C$.

\subsection{Specific Heat at Fixed Particle Number Density}
The standard description of the BEC phase transition in a
non-relativistic Bose gas is discussed in terms of the specific
heat per particle at finite volume, $C_V/N$ (see, e.g.,
Refs.~\cite{Landau,Tolmachjev,Rumer,Greiner}). The relativistic
analog of this quantity is:
\eq{ \label{CVTL}
\frac{c_V}{\rho}
 \;\equiv\; \frac{1}{\rho}\;\left(\frac{\partial \varepsilon}{\partial T}
 \right)_{\rho}~.%\label{CV}
}
The energy density in the TL equals to:
 \eq{\label{epsTc}
 &\varepsilon(T,\mu)~=~\rho_C~m~+~\varepsilon^*(T,\mu)~=~
 \rho_C~m~+~\frac{3}{2\pi^2}\int_{0}^{\infty}p^2dp~\frac{\sqrt{m^2+p^2}}
 {\exp\left[\left(\sqrt{m^2+p^2}~-~\mu\right)/T\right]-1}\;\nonumber \\
 &= \rho_C~m~+~\frac{3T^2m^2}{2\pi^2}\sum_{n=1}^{\infty}\left\{\frac{1}{n^2}
 K_2\left(\frac{nm}{T}\right)
 + \frac{m}{2nT} \left[K_1\left(\frac{nm}{T}\right) + K_3\left(\frac{nm}{T}
 \right)\right]
 \right\}\exp\left(\frac{n\mu}{T}\right)~,
}
where $\rho_C=0$ and $\mu\le m$ at $T\ge T_C$, while $\rho_C>0$
and $\mu=m$ at $T<T_C$.
The high and low temperature behavior of $c_V/\rho$ can be easily
found. At $T\rightarrow\infty$ and fixed $\rho$, both the Bose
effects and particle mass become inessential. The energy density
(\ref{epsTc}) behaves as, $\varepsilon \cong 3T\rho$, thus,
$c_V/\rho\cong 3$. Note that in a non-relativistic gas at $T_C\ll
T\ll m$, one finds $\varepsilon \cong (m +3T/2)\rho$. Thus, the
`high-temperature non-relativistic limit' would give
$c_V/\rho\cong 3/2$.
At $T_C\gg T\rightarrow 0$ the behavior of $\varepsilon$ at fixed
total particle number density, $\rho=\rho_C+\rho^*(T,\mu=m)$, is
given by,
\eq{\label{eps-nonrel-2}
 \varepsilon~=~
 \rho~ m~
 +~\frac{9~\zeta(5/2)}{2}\left(\frac{m}{2\pi}\right)^{3/2}~T^{5/2}~.
 }
 This leads to:
\eq{\label{CV-T0}
\left(\frac{c_V}{\rho}\right)_{T\rightarrow
0}~\cong~\frac{45~\zeta(5/2)}{4~\rho}~\left(\frac{m}{2\pi}\right)^{3/2}~T^{3/2}~\propto~T^{3/2}~\rightarrow
0~.
}
\begin{figure}[h!]
 \epsfig{file=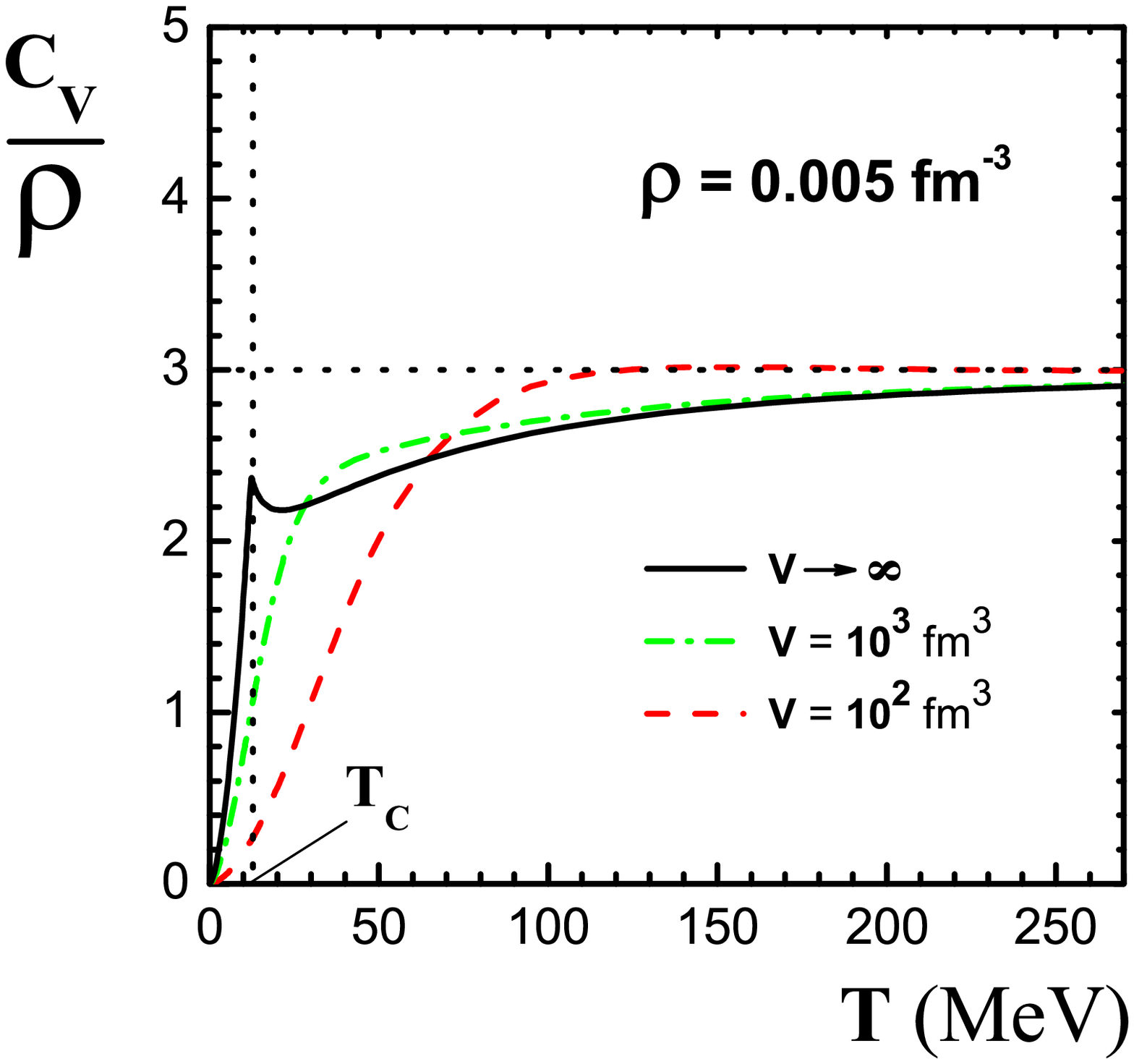,width=0.49\textwidth}\;
 \epsfig{file=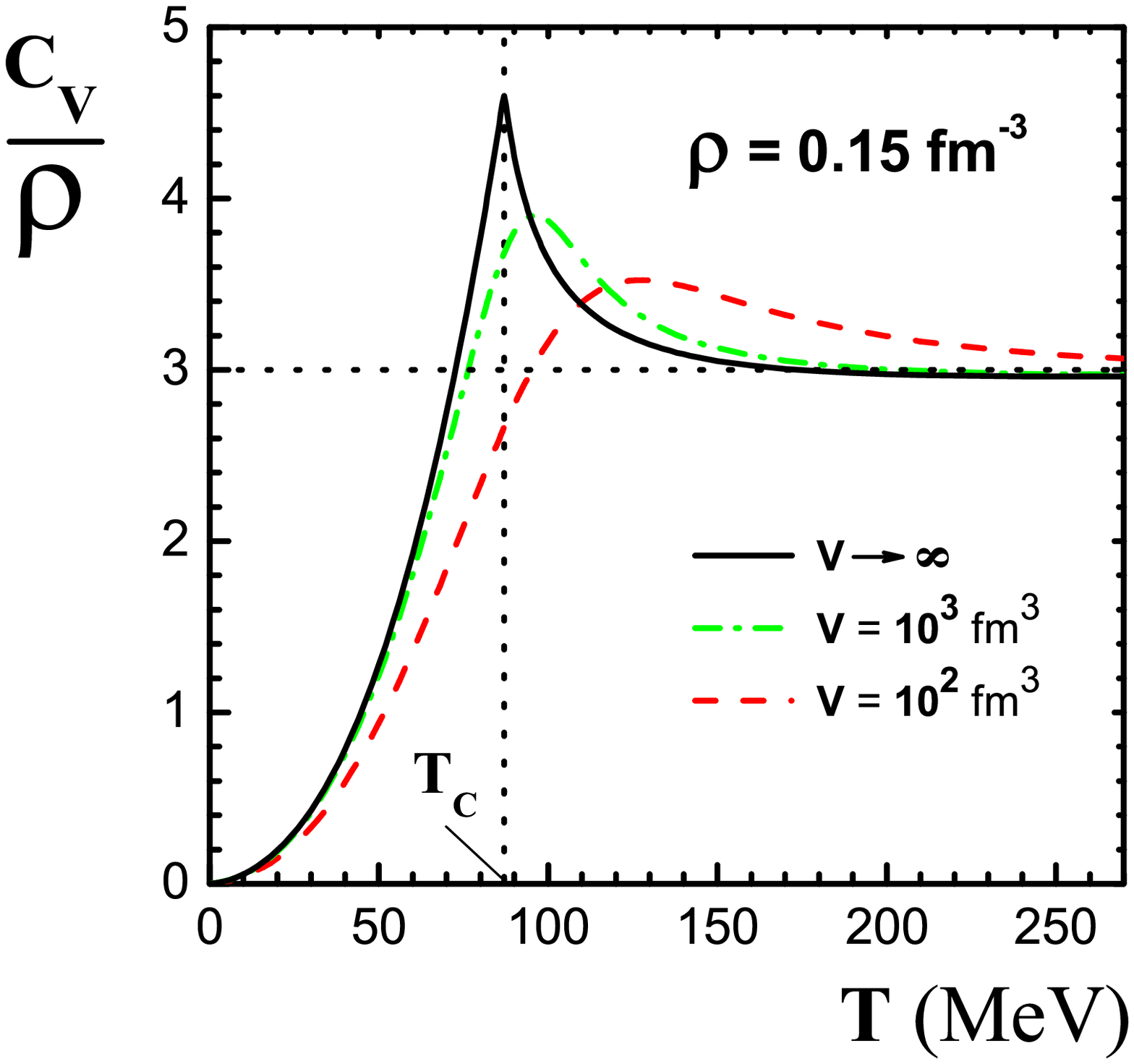,width=0.49\textwidth}
\caption{The solid lines demonstrate the temperature dependence of
$c_V/\rho$ (\ref{CVTL}) in the TL at fixed values of $\rho=0.005$
and 0.15~fm$^{-3}$. The $c_V/\rho$ in the finite system is
described by Eq.~(\ref{CV-V}) (see next Section).  The dashed
lines correspond to $V=10^2$~fm$^3$ and the dashed-dotted lines to
$V=10^3$~fm$^3$, respectively.}
 \label{fig-Cv}
\end{figure}

The Fig.~\ref{fig-Cv} shows the temperature dependence of
$c_V/\rho$ at fixed $\rho$.
As seen from Fig.~\ref{fig-Cv}, $c_V/\rho$ (\ref{CVTL}) has a
maximum at $T=T_C$. The $c_V/\rho$ is a continuous function of
$T$, whereas its temperature derivative has a discontinuity at
$T=T_C$. This discontinuity emerges in the TL $V\rightarrow\infty$
and can be classified as a 3rd order phase transition. To estimate
the value of $c_V/\rho$ (\ref{CVTL}) at $T=T_C$ we start from
$T>T_C$ when the contribution from $p=0$ level to $c_V/\rho$
(\ref{CVTL}) equals to zero in the TL, and then consider the limit
$T\rightarrow T_C+0$. We discuss separately the non-relativistic
and ultra-relativistic approximations.

Using the asymptotic, $K_{\nu}(x)\cong
\sqrt{\pi/(2x)}\exp(-x)[1+(4\nu^2-1)/8x]~$ at $x\gg 1$ \cite{Abr},
one finds a non-relativistic limit of Eqs.~(\ref{rho-pi}) and
(\ref{epsTc}), respectively,
\eq{\label{rho-nonrel}
& \rho(T,\mu)~\cong~ 3\left(\frac{m ~T} {2\pi}\right)^{3/2}~\left[
Li_{3/2}(z)~+~\frac{15~T}{8~m}~Li_{5/2}(z)\right]~,\\
& \varepsilon(T,\mu)
 =
 3\left(\frac{mT}{2\pi}\right)^{3/2}m
 \left[Li_{3/2}(z)+\frac{27T}{8m}Li_{5/2}(z)\right]=
 \rho m +3\left(\frac{m}{2\pi}\right)^{3/2}T^{5/2}\frac{3}{2}Li_{5/2}(z)~.\label{eps-nonrel}
 }
 where $z\equiv \exp[-(m-\mu)/T]$ and
$Li_k(z)=\sum_{k=1}^{\infty}z^n/n^k$ is the polylogarithm function
\cite{Prud}.
%From Eq.~(\ref{epsTc}) one finds,
%The condition $d\rho/dT=0$ used to
%Eq.~(\ref{rho-nonrel}) gives
%
%\eq{\label{mu-prime}
%
%\frac{m~-\mu}{T}~+\mu^{\prime}~\cong~-~\left[\frac{3}{2}~Li_{3/2}~+~
%\frac{75T}{16m}~{Li_{5/2}(z)}\right]~\times~
%\left[Li_{1/2}(z)~+~\frac{15T}{8m}~Li_{3/2}(z)\right]^{-1}~.
%
%}
%
%The momentum integral in Eq.~(\ref{CV}) in a non-relativistic
%limit can be calculated as
%
%\eq{\label{CV-nonrel}
%
%
%
% For the derivative $d\varepsilon/dT$ one finds from
% Eq.~(\ref{epsTc}),
% \eq{
% \left[3\left(\frac{mT}{2\pi}\right)^{3/2}\right]^{-1}~ \frac{d\varepsilon}{d
% T}~&\cong~ \frac{135}{16}~Li_{5/2}(z)~+~
% \frac{3m}{2T}~Li_{3/2}(z)~
%+~\frac{95}{8}~Li_{5/2}(z)
% \nonumber \\
% &+~\left[\frac{m}{T}~Li_{1/2}(z)~+~\frac{27}{8}~Li_{3/2}(z)\right]
% \left(\frac{m~-~\mu}{T}~+~\mu^{\prime}\right)
%
%~. \label{epsilon-prime}
% }
%
At $T=T_C$ it follows, $z=1$ and $Li_k(1)=\zeta(k)$. Using
Eqs.~(\ref{rho-nonrel}-\ref{eps-nonrel}), and (\ref{A4}) one finds
at $T=T_C\ll m$,
\eq{\label{CV1}
\left(\frac{c_V}{\rho}\right)_{T=T_C}~\cong
~\frac{15}{4}~\frac{\zeta(5/2)}{\zeta(3/2)}~\cong~ 1.926~.
}
%
%
%, and
%$\zeta(1/2)=\infty $.  This gives, at $m\gg T=T_C$,
%
%\eq{\label{CVTC}
%
%\left(\frac{c_V}{\rho}\right)_{T_{C}}~\cong
%~\frac{15}{14}~\frac{\zeta(5/2)}{\zeta(3/2)}~\cong~ 1.925~.
%
%}
%
%The `high temperature' non-relativistic limit, $m\gg T\gg T_C$,
%leads to $z\rightarrow 0$ and $Li_k(z)\rightarrow 1$. The
%Eq.~(\ref{CV1}) then gives,
%
%\eq{\label{CVT}
%
%\frac{c_V}{\rho}~\cong~\frac{15}{4}~-~\frac{9}{4}~=~\frac{3}{2}~.
%
%}
%
%
%Using the asymptotic expansion, $K_2(x)\cong 2/x^2$ at $x\ll 1$,
%one finds the ultra-relativistic limit $T>>m$ from (\ref{rho-pi})
%at $T>T_C$,
%
%\eq{\label{rho-rel}
%
%\rho(T,\mu)~\cong~\frac{3T^3}{\pi^2}~\sum_{n=1}^{\infty}\frac{1}{n^3}
%\left(1~+~ \frac{n~\mu
%}{T}\right)~=~\frac{3T^3}{\pi^2}~\zeta(3)~+~\frac{3T^2}{\pi^2}~\zeta(2)~\mu~.
%
%}
%
% The condition $d\rho/dT=0$ used to Eq.~(\ref{rho-rel}) gives
%
%\eq{\label{mu-prime}
%
%\mu^{\prime}~\cong~-~3~\frac{\zeta(3)}{\zeta(2)}~.
%
%}
%
Using the asymptotic expansion,
$K_{\nu}(x)\cong\frac{1}{2}\Gamma(\nu)(x/2)^{-\nu}$ at $x\ll 1$
\cite{Abr}, one finds from Eq.~(\ref{epsTc}) in the
ultra-relativistic limit $T\ge T_C\gg m$,
\eq{\label{eps-rel}
\varepsilon(T,\mu)~\cong~\frac{3T^4}{\pi^2}~\sum_{n=1}^{\infty}\left[\frac{3}{n^4}~\left(1~+
~\frac{n\mu}{T}\right)\right]~=~\frac{9\zeta(4)}{\pi^2}~T^4~+~\frac{9\zeta(3)}{\pi^2}~T^3~\mu~.
}
% & \frac{d\varepsilon}{dT}~\cong ~\frac{9T^3}{\pi^2}~\left[4
%\zeta(4)~-~3~\frac{\zeta(3)\zeta(3)}{\zeta(2)}\right]~.
%
%}
%
From Eqs.~(\ref{eps-rel}) and (\ref{TC-rel}) it then follows at
$m\ll T = T_C$,
\eq{\label{CV2}
\left(\frac{c_V}{\rho}\right)_{T=T_C}~\cong~12~\frac{\zeta(4)}{\zeta(3)}~\cong~
10.805~.
}
Different pion number densities correspond to different values of
the BEC temperature $T_C$. The Eqs.~(\ref{CV1},\ref{CV2})  show
that $c_V/\rho$  goes at $T\rightarrow\infty$ to its limiting
value $3$ from below, if $T_C$ is `small', and from above, if
$T_C$ is `large' (see Fig.~\ref{fig-Cv}).

Note that at the BEC in atomic gases the number of atoms is
conserved. Thus, the temperature dependence of $c_V/\rho$ for the
system of atoms at fixed $\rho$ can be straightforwardly measured.
This is much more difficult for the pion gas. There is no
conservation law of the number of pions, and the special
experimental procedure is needed to form the statistical ensemble
with fixed number of pions.

% \eq{
% K_1(x) \;\simeq\; 1/x\;,\quad
% K_2(x) \;\simeq\; 2/x^2\;,\quad
% K_3(x) \;\simeq\; 8/x^3 \;-\; 1/x\;.
% }
%
% \eq{
% \sum_{n=1}^{\infty} \frac{1}{n^k} \;=\; \zeta(k)\;.
% }
%
% \eq{
% & \zeta(4) \;=\; \pi^4/90 \;\simeq\; 1.1\;,\quad
% \zeta(3) \;\simeq\; 1.2\;,\quad
% 12 \zeta(4)/\zeta(3) \;\simeq\; 10.8\;.\nonumber \\
% & \zeta(4)=1.0823~,\quad \zeta(3)= 1.202~,\quad
% \zeta(5/2)=1.341~,\quad \zeta(3/2)=2.612\nonumber \\
% & \zeta(5/2)/\zeta(3/2)= 0.5134~,\quad \zeta(2)=1.645
% }
%
%\eq{K_{\nu}(z)\cong\sqrt {\frac{\pi}{2z}}~\exp(-z)~(1~+~
%\frac{1}{z}~\frac{4\nu^2~-~1}{8})
%
%}
%

%

%
%%%%%%%%%%%%%%%%%%%%%%%%%%%%%%%%%%%%%%%%%%%%%%%%%%%%%%%%%%%%%%%%%%%%%%%%%%%%%%%%%%%%%%%%%%%%%%
%
\section{Finite Size Effects}\label{s-FSE}
%
%%%%%%%%%%%%%%%%%%%%%%%%%%%%%%%%%%%%%%%%%%%%%%%%%%%%%%%%%%%%%%%%%%%%%%%%%%%%%%%%%%%%%%%%%%%%%%
%
\subsection{Chemical Potential at Finite Volume}
The standard introduction of $\rho_C$ with Eq.~(\ref{rho-cond}) is
rather formal. To have a more realistic picture, one needs to
start with finite volume system and consider the limit
$V\rightarrow \infty$ explicitly. The main problem is that the
substitution, $\sum_{\bf{p}}\ldots \cong
(V/2\pi^2)\int_0^{\infty}\ldots p^2dp$, becomes invalid below the
BEC line.  We consider separately the contribution to the total
pion density from the two lower quantum states,
 \eq{\label{Np0}
 \rho &~\cong ~\frac{1}{V}~\sum_{{\bf p},j}^{\infty} \langle n_{{\bf p},j} \rangle
% \;=\; \rho_C \;+\; \rho_1 \;+\; \rho^*_{\bf p>p_1} \nonumber
% \\
 ~=~
 \frac{3}{V}~\frac{1}{\exp\left[\left(m-\mu\right)/T\right]\;-\;1}
 \;+\;\frac{3}{V}~\frac{6}{\exp\left[\left(\sqrt{m^2+p_1^2}-\mu\right)/T\right]-1}
 \nonumber\\
 &+\; \frac{3}{2\pi^2}\int_{p_1}^{\infty}p^2dp~\frac{1}
 {\exp\left[\left(\sqrt{m^2+p^2}-\mu\right)/T\right]-1}~.
 }
The first term in the r.h.s. of Eq.~(\ref{Np0}) corresponds to the
lowest momentum level $p=0$,  the second one to the  first excited
level $p_1=2\pi V^{-1/3}$ with the degeneracy factor 6, and the
third term approximates the contribution from levels with $p>
p_1=2\pi V^{-1/3}$. Note that this corresponds to free particles
in a box with periodic boundary conditions (see, e.g.,
Ref~\cite{Tolpygo}). At any finite $V$ the equality $\mu=m$ is
forbidden as it would lead to the infinite value of particle
number density at $p=0$ level.

At $T<T_C$ in the TL  one expects a finite non-zero particle
density $\rho_C$ at the $p=0$ level. This requires
$(m-\mu)/T\equiv \delta\propto V^{-1}$ at $V\rightarrow \infty$.
The particle number density at the $p=p_1$ level can be then
estimated as,
\eq{
\rho_1=\frac{18~V^{-1}}{\exp\left[\left(\sqrt{m^2+p_1^2}-\mu\right)/T\right]-1}
\cong
\frac{18~V^{-1}}{\delta~+~p^2_1/(2mT)}~\propto~\frac{V^{-1}}{V^{-2/3}}~=~V^{-1/3}~,
}
and it goes to zero at $V\rightarrow\infty$. Thus, the second term
in the r.h.s. of Eq.~(\ref{Np0}) can be neglected in the TL. One
can also extend the lower limit of integration in the third term
in the r.h.s. of Eq.~(\ref{Np0}) to $p=0$, as the region $[0,p_1]$
contributes as $V^{-1/3}\rightarrow 0$ in the TL and can be safely
neglected. Therefore, we consider the pion number density and
energy density at large but finite $V$ in the following form:
 \eq{\label{rhoV}
 \rho &~\cong ~
 \frac{3}{V}~\frac{1}{\exp\left[\left(m-\mu\right)/T\right]\;-\;1}
 \;+\;\rho^*(T,\mu)~,
 % \frac{3}{2\pi^2}\int_{0}^{\infty}p^2dp~\frac{1}
 %{\exp\left[\left(\sqrt{m^2+p^2}-\mu\right)/T\right]-1}\;,
 \\
 \varepsilon &
 %~= ~ \frac{1}{V}~\sum_{{\bf p},j}^{\infty}
 %\sqrt{m_j^2+p^2}\cdot \langle n_{{\bf p},j} \rangle
 \;\cong~
 \frac{3}{V}~\frac{m}{\exp\left[\left(m-\mu\right)/T\right]\;-\;1}~
+\; \varepsilon^*(T,\mu)~.\label{epsV}
%\frac{3}{2\pi^2}\int_{0}^{\infty}p^2dp~\frac{\sqrt{m^2+p^2}}
% {\exp\left[\left(\sqrt{m^2+p^2}-\mu\right)/T\right]-1}\;.\label{epsV}
% \equiv~\varepsilon_C \;+\; \varepsilon^* \nonumber.
 %
 }
Thus, at large $V$, the zero momentum level defines completely the
finite size effects of the pion system.

\begin{figure}[h!]
 \epsfig{file=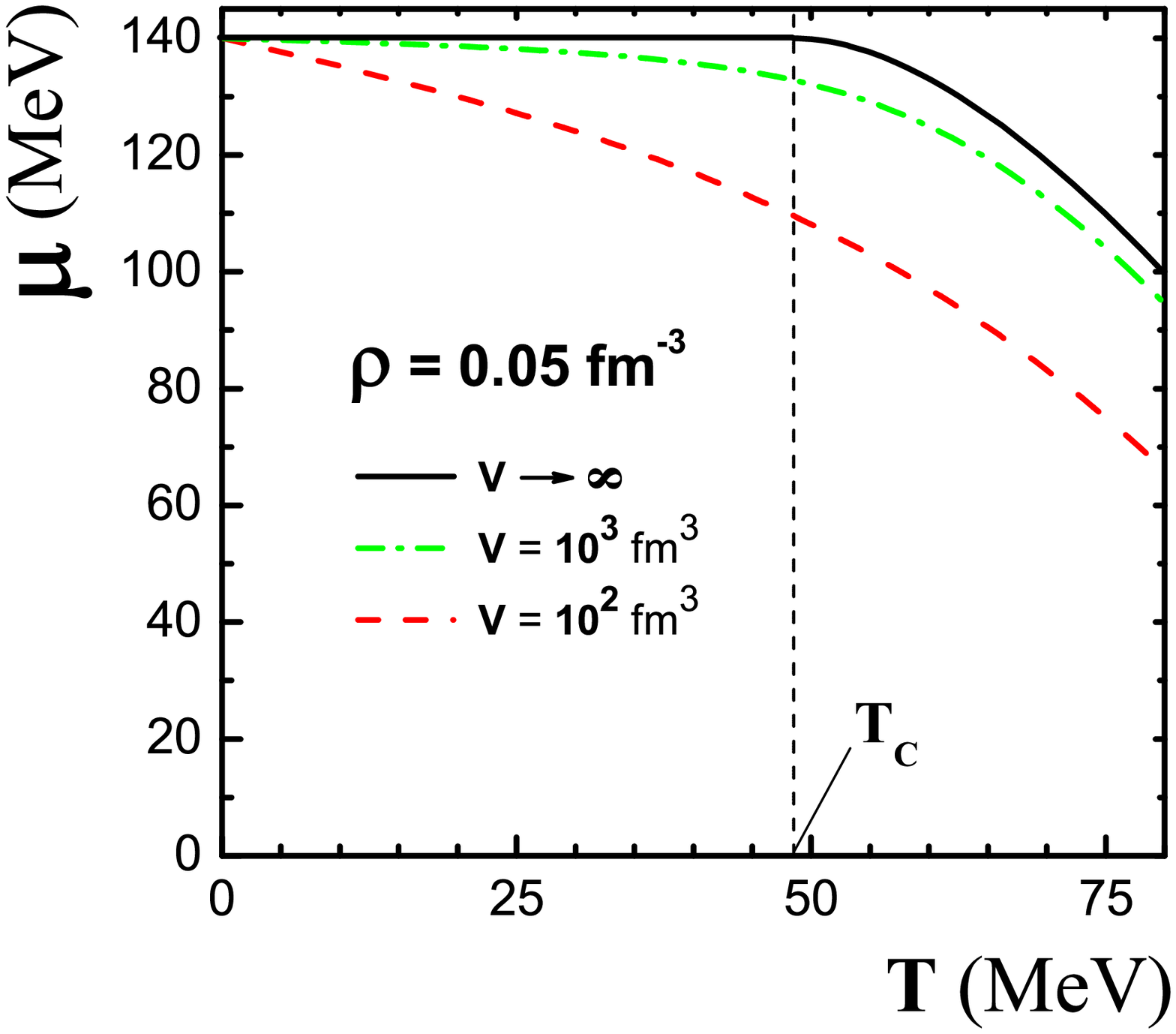,width=0.49\textwidth}\;
 \epsfig{file=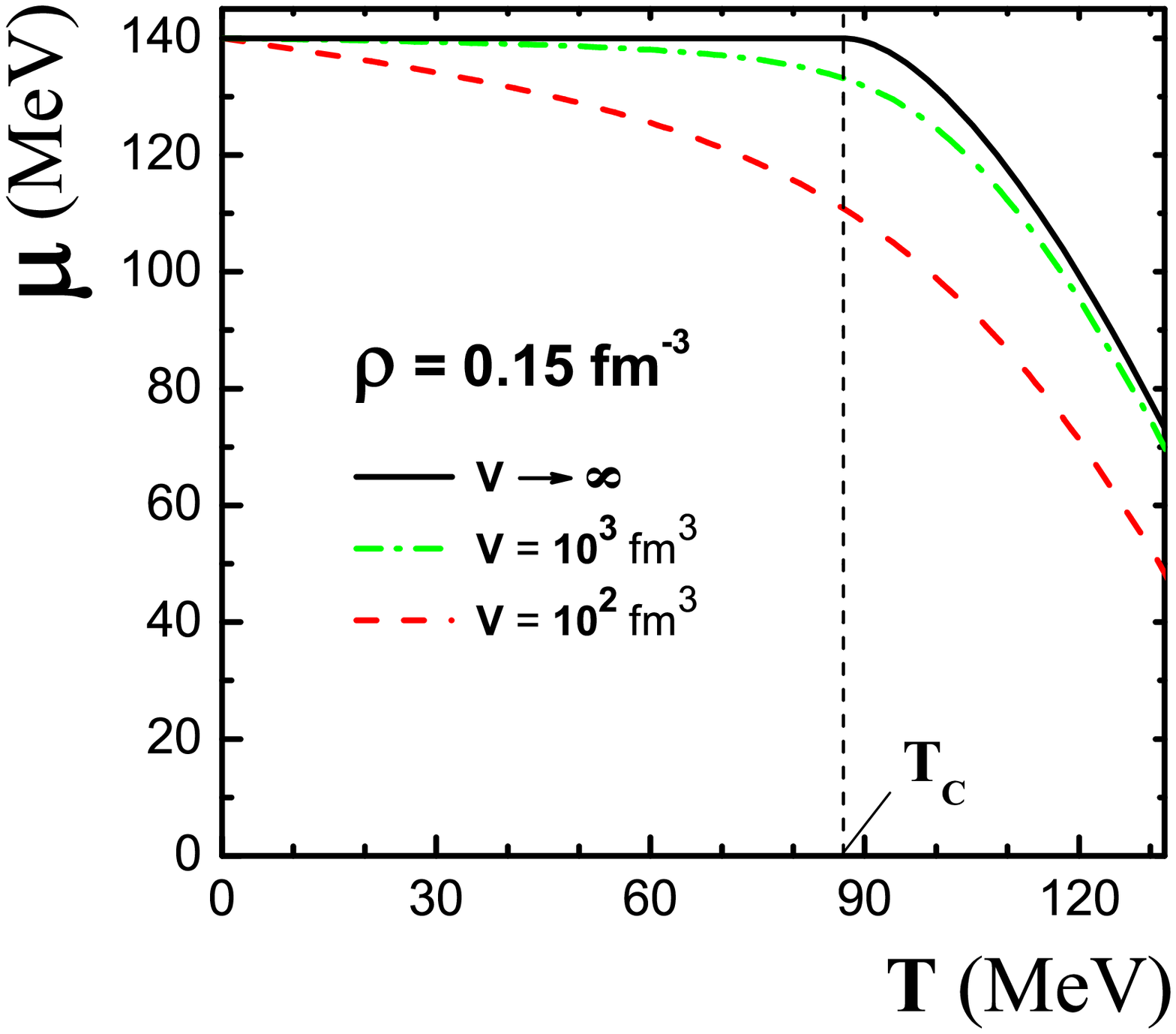,width=0.49\textwidth}
\caption{The chemical potential $\mu$ as a function of temperature
$T$ at fixed particle number density $\rho$. The solid line
presents the behavior in the TL $V\rightarrow\infty$. The dashed
line corresponds to $V=10^2$~fm$^3$, and dashed-dotted line to
$V=10^3$~fm$^3$. The vertical dotted line indicates the BEC
temperature $T_C$. The left panel corresponds to
$\rho=0.05$~fm$^{-3}$, the right one to $\rho=0.15$~fm$^{-3}$. }
 \label{fig-mu}
\end{figure}

The behavior of $\rho^*(T,\mu)$ at $\mu \rightarrow m$ can be
found from Eq.~(\ref{A1}).
%
%\eq{\label{delta1}
%
%\Delta ~\equiv~\rho^*(T,\mu=m)~-~\rho^*(T,\mu)~\cong ~
%\frac{3}{\sqrt{2}\pi}~(m T)^{3/2}~\sqrt{\delta}~.
 %
% }
%
%
%The numerical results for $\Delta $ at different values of $T$ are
%shown in Fig.~\ref{fig-delta} by the solid lines, while the dashed
%lines present the asymptotic behavior of the r.h.s. of
%Eq.~(\ref{delta1}).
%
%\begin{figure}[h!]
% \epsfig{file=fig2.eps,width=0.6\textwidth}
% \caption{
%$\Delta \equiv \rho^*(T,\mu=m)-\rho^*(T,\mu)$ as a function of
%$\delta=(m-\mu)/T$ at different temperatures, $T=10,~50,~100$, and
%$150$~MeV. The solid lines present the numerical results, whereas
%the dashed lines indicate the asymptotic expansion at
%$\delta\rightarrow 0$ according to Eq.~(\ref{delta1}).
% }
% \label{fig-delta}
%\end{figure}
%
%
%Finally,
At large $V$, Eq.~(\ref{rhoV}) takes then the following form,
 \eq{\label{N}
 \rho\;\cong \;
 \frac{3}{V~\delta}~+~\rho^*(T,\mu)~\cong~\frac{3}{V~\delta}~+~\rho^*(T,\mu=m)~
 -~\frac{3}{\sqrt{2}\pi}~(m T)^{3/2}~\sqrt{\delta}~.
 }
The Eq.~(\ref{N}) can be written as,
 \eq{\label{Eq}
 A\,\delta^{3/2} \;+\; B\,\delta \;-\;1 \;=\; 0\;,
 }
where
 \eq{\label{A}
  A &\;=\;\frac{V}{\sqrt{2}\pi}~(m T)^{3/2}~\equiv~ a(T)~V~,
  %\frac{2\sqrt{\pi}}{\zeta[3/2]}\; g^{-1}\langle N_{tot}\rangle
 % \left(\frac{T}{T_C}\right)^{3/2}
 %   \;\simeq\; 1.4\,g^{-1}\langle N_{tot}\rangle\, \left(T/T_C\right)^{3/2}\;,
 %   \nonumber
 \\
 B &\;=~\frac{V}{3}~[~\rho~-~\rho^*(T,\mu=m)~]~\equiv~b(T)~V~.
 %\;g^{-1}\langle N_{tot}\rangle\,\left[1 \;-\; \left(\frac{T}{T_C}\right)^{3/2}\right]\;.
 \label{B}
 }
The Eq.~(\ref{Eq}) for $\delta$ has two complex roots and one real
root. An asymptotic behavior at $V\rightarrow\infty$ of the
physical (real) root can be easily found. At $T<T_C$ it follows
from Eq.~(\ref{B}) that $b(T)>0$, and one finds from
Eq.~(\ref{Eq}) at large $V$,
\eq{\label{deltaT1}
\delta~\cong~\frac{1}{b}~V^{-1}~.
}
From Eq.~(\ref{B}) one finds that $b=0$ at $T=T_C$. In this case,
Eq.~(\ref{Eq}) gives,
\eq{\label{deltaT2}
\delta~\cong~\frac{1}{a^{2/3}}~V^{-2/3}~.
}
The Eq.~(\ref{Eq}) can be also used at $T>T_C$, if $T$ is close to
$T_C$, thus, $\delta \ll 1$. In this case it follows from
Eq.~(\ref{B}) that $b(T)<0$, and one finds from Eq.~(\ref{Eq}),
\eq{\label{deltaT3}
\delta~\cong~\frac{b^2}{a^2}~.
}
Thus, $\delta$ is small but finite at $V\rightarrow\infty$, and
$\mu$ remains smaller than $m$ in the TL. The temperature
dependence of chemical potential $\mu=\mu(T)$ for $V=10^2$~fm$^3$
and $10^3$~fm$^3$ at fixed pion number density $\rho$ is shown in
Fig.~\ref{fig-mu}.
%two different
%particle number densities $\rho=0.05, 0.15 fm^{-3}$ and
%

The value of $c_V/\rho$ (\ref{CVTL}) at finite volume $V$ is
calculated as,
 \eq{\label{CV-V}
 & \frac{c_V}{\rho}
 \;\equiv\; \frac{1}{\rho}\;\left(\frac{\partial \varepsilon}{\partial T} \right)_{\rho,V}
 %\nonumber
 %\\
 \;=\; \frac{3}{\rho\,V}~\frac{\exp\left[\left(m-\mu\right)/T\right]}
 {\left(\exp\left[\left(m-\mu\right)/T\right]\;-\;1\right)^2}~
 \frac{m\left(m-\mu+\mu'\,T\right)}{T^2}
  \\
 &\;+\; \frac{3}{2\pi^2\rho}\int_{0}^{\infty}p^2dp~
 \frac{\exp\left[\left(\sqrt{m^2+p^2}-\mu\right)/T\right]}
 {\left(\exp\left[\left(\sqrt{m^2+p^2}-\mu\right)/T\right]-1\right)^2}\;
 \frac{\sqrt{m^2+p^2}\left(\sqrt{m^2+p^2}-\mu+\mu'\,T\right)}{T^2}\;,
 \nonumber
 }
where $\mu'=(\partial \mu/\partial T)_{\rho,V}$.  The temperature
dependence of $c_V/\rho$ (\ref{CV-V}) at several fixed values of
$V$ is shown in Fig.~\ref{fig-Cv} by the dashed and dashed-dotted
lines.

\subsection{Particle Number Fluctuations}\label{s-Mult-Fluct}
\begin{figure}[h!]
 \epsfig{file=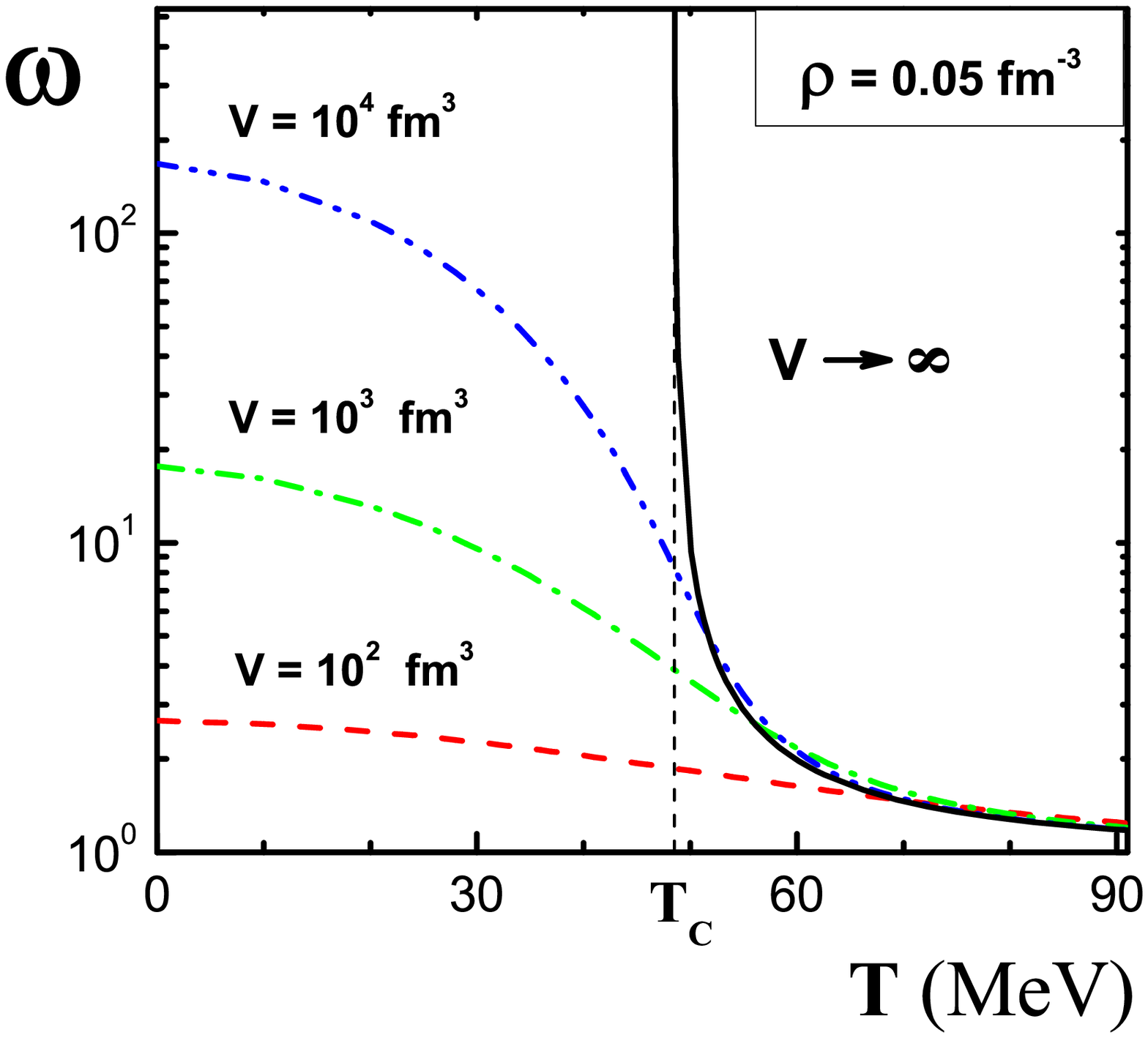,width=0.49\textwidth}
 \epsfig{file=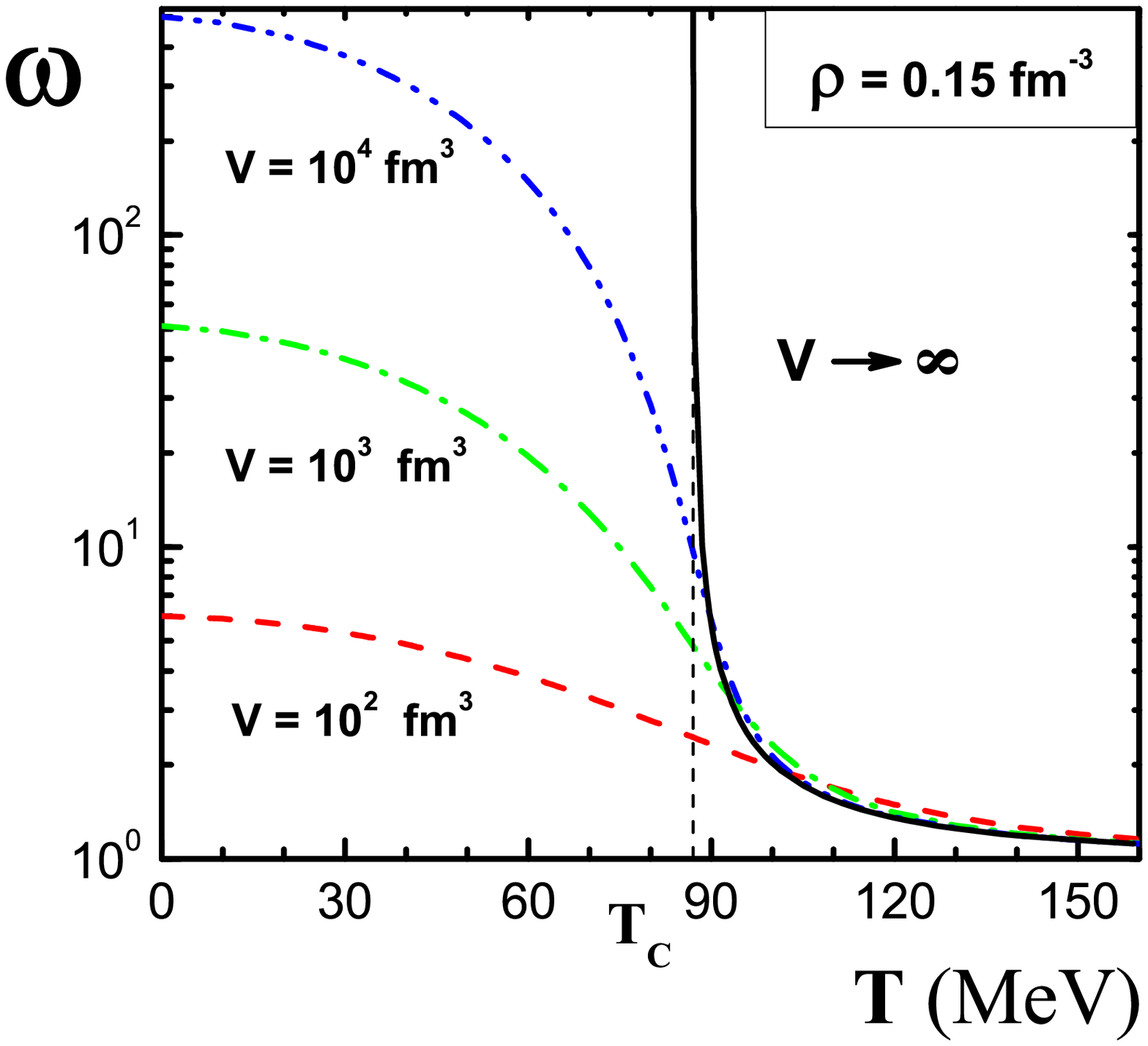,width=0.49\textwidth}
\caption{The dashed lines show the GCE scaled variance
(\ref{W_def}) for the pion gas as a function of temperature $T$
for $V=10^4$~fm$^3$, $10^3$~fm$^3$, $10^2$~fm$^3$ (from top to
bottom). The vertical dotted line indicates the BEC temperature
$T_C$. The solid line shows the $\omega$ (\ref{W_def}) in the TL
$V\rightarrow\infty$. The left panel corresponds to
$\rho=0.05$~fm$^{-3}$, the right one to $\rho=0.15$~fm$^{-3}$~. }
 \label{fig-W0gce}
\end{figure}

%
%
%Similarly to the procedure discussed discussed above in this
%section one can find for the variance of
The variance of particle number fluctuations in the GCE at finite
$V$ is:
 \eq{\label{DN2}
& \langle\Delta N^2\rangle ~\equiv~ \langle \left(N~-~\langle
N\rangle\right)^2\rangle
 ~=~\sum_{{\bf p},j}\langle n_{{\bf p},j}\rangle(1+\langle n_{{\bf p},j}\rangle)
\;\cong \; \frac{3}{\exp\left[\left(m-\mu\right)/T\right]\;-\;1}
\nonumber
 \\
&   \;+\;
\frac{3}{\left\{\exp\left[\left(m-\mu\right)/T\right]\;-\;1\right\}^2}
% \nonumber
% \\
 \;+\; \frac{3V}{2\pi^2}\int_{0}^{\infty}p^2dp\;
        \frac{\exp\left[\left(\sqrt{m^2+p^2}-\mu\right)/T\right]}
        {\left\{\exp\left[\left(\sqrt{m^2+p^2}-\mu\right)/T\right]\;-\;1\right\}^2}~~,
}
where the first two terms in the r.h.s. of Eq.~(\ref{DN2})
correspond to particles of the lowest level $p=0$, and the third
term to particles with $p>0$.
%The degeneracy factor $g$ equal to
%unity in the Eq.~(\ref{<DN2>}), because we consider fluctuations
%of $\pi^0$ mesons in the $\pi^+$, $\pi^-$, $\pi^0$ gas.
%
We will use the scaled variance,
 \eq{\label{W_def}
 \omega \;=\; \frac{\langle\Delta N^2\rangle}{\langle N\rangle}\;,
 }
as the measure of particle number fluctuations. The numerical
results for the scaled variance are shown in Fig.~\ref{fig-W0gce}.
At $T>T_C$ the parameter $\delta$ goes to the finite limit
(\ref{deltaT3}) at $V\rightarrow\infty$. This leads to the finite
value of $\omega$ (\ref{W_def}) in the TL.
At $T\le T_C$ one finds from Eq.~(\ref{DN2}) in the TL,
 \eq{\label{DN2a}
 \langle\Delta N^2\rangle
 \;\cong\; 3~\delta^{-2}
  \;+\; \frac{3}{2}~V~a~\delta^{-1/2}\;.
 }
This gives,
 \eq{\label{Wpi0}
 \omega \;\equiv\; \frac{\langle\Delta N^2\rangle}{\langle N \rangle}
 \;\cong~ 3~\rho^{-1}~V^{-1}~\delta^{-2}
 \;+\;
 \frac{3}{2}~a~\rho^{-1}~\delta^{-1/2}\,,
 }
where $a=a(T)$ is defined in Eq.~(\ref{A}).  The substitution of
$\delta$ in Eq.~(\ref{Wpi0}) from (\ref{deltaT1}), gives for
$T<T_C$ and $V\rightarrow\infty$,
 \eq{\label{omegaT1}
 \omega~\cong~ 3~b^2~\rho^{-1}~V~+~
 \frac{3}{2}~a~b^{1/2}~\rho^{-1}~V^{1/2}~\equiv~\omega_C~+~\omega^*~.
 }
The $\omega_C$ in the r.h.s. of Eq.~(\ref{omegaT1}) is
proportional to $V$ and corresponds to the particle number
fluctuations in the BE condensate, i.e. at the $p=0$ level,
$\omega_C\cong \sum_{j}\langle(\Delta n_{p=0,j})^2\rangle/\langle
N\rangle$. The second term, $\omega^*$ is proportional to
$V^{1/2}$. It comes from the fluctuation of particle numbers at
$p>0$ levels, $\omega^*=\sum_{{\bf p},j; p>0}\langle (\Delta
n_{{\bf p},j})^2\rangle/\langle N\rangle$.  At $T\rightarrow 0$,
one finds $a\rightarrow 0$ and $b\rightarrow \rho/3$. This gives
the maximal value of the scaled variance, $\omega =\rho
V/3=\langle N\rangle/3$, for given $\rho$ and $V$ values.

\begin{figure}[h!]
 \epsfig{file=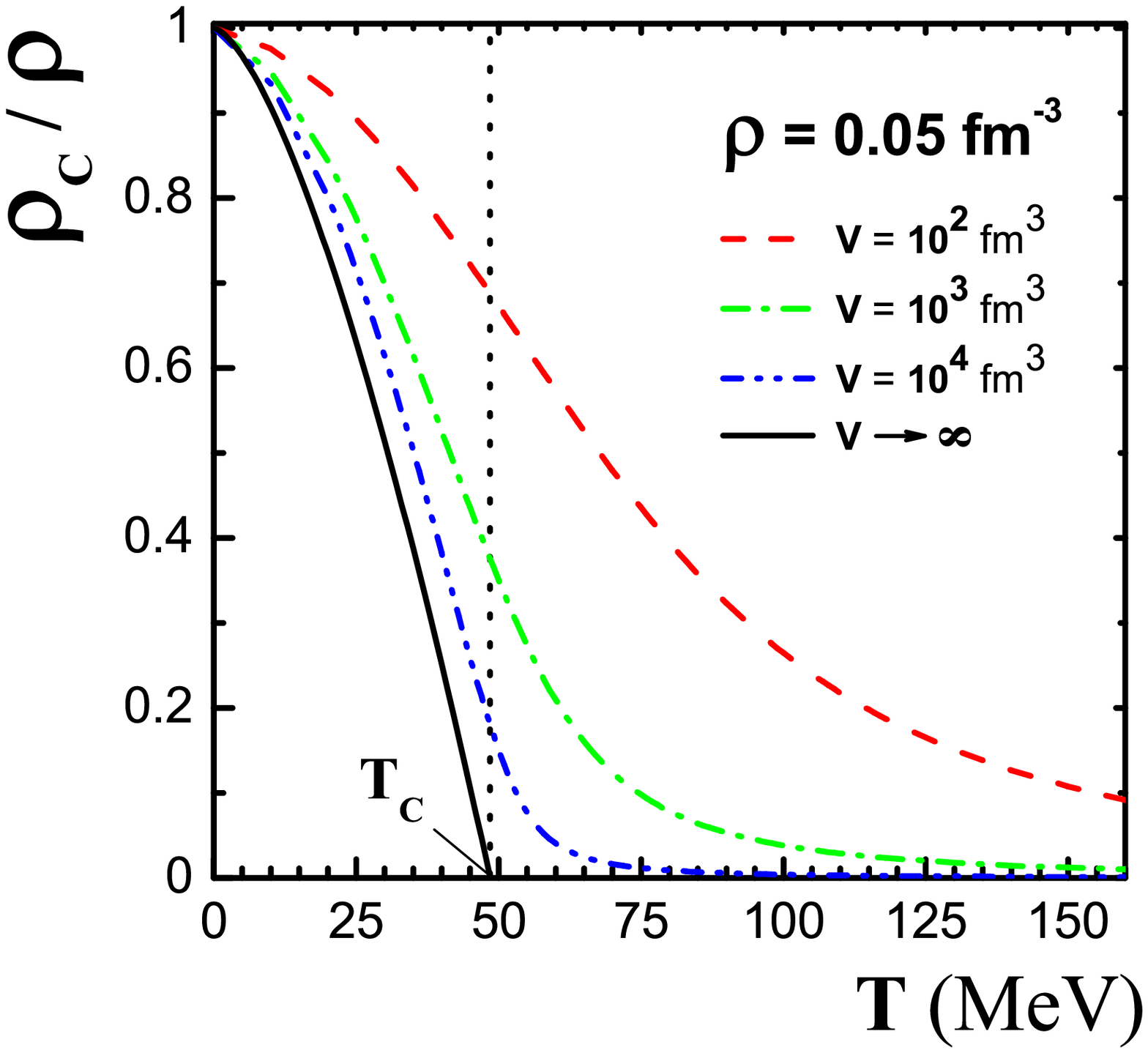,width=0.45\textwidth}\quad
 \epsfig{file=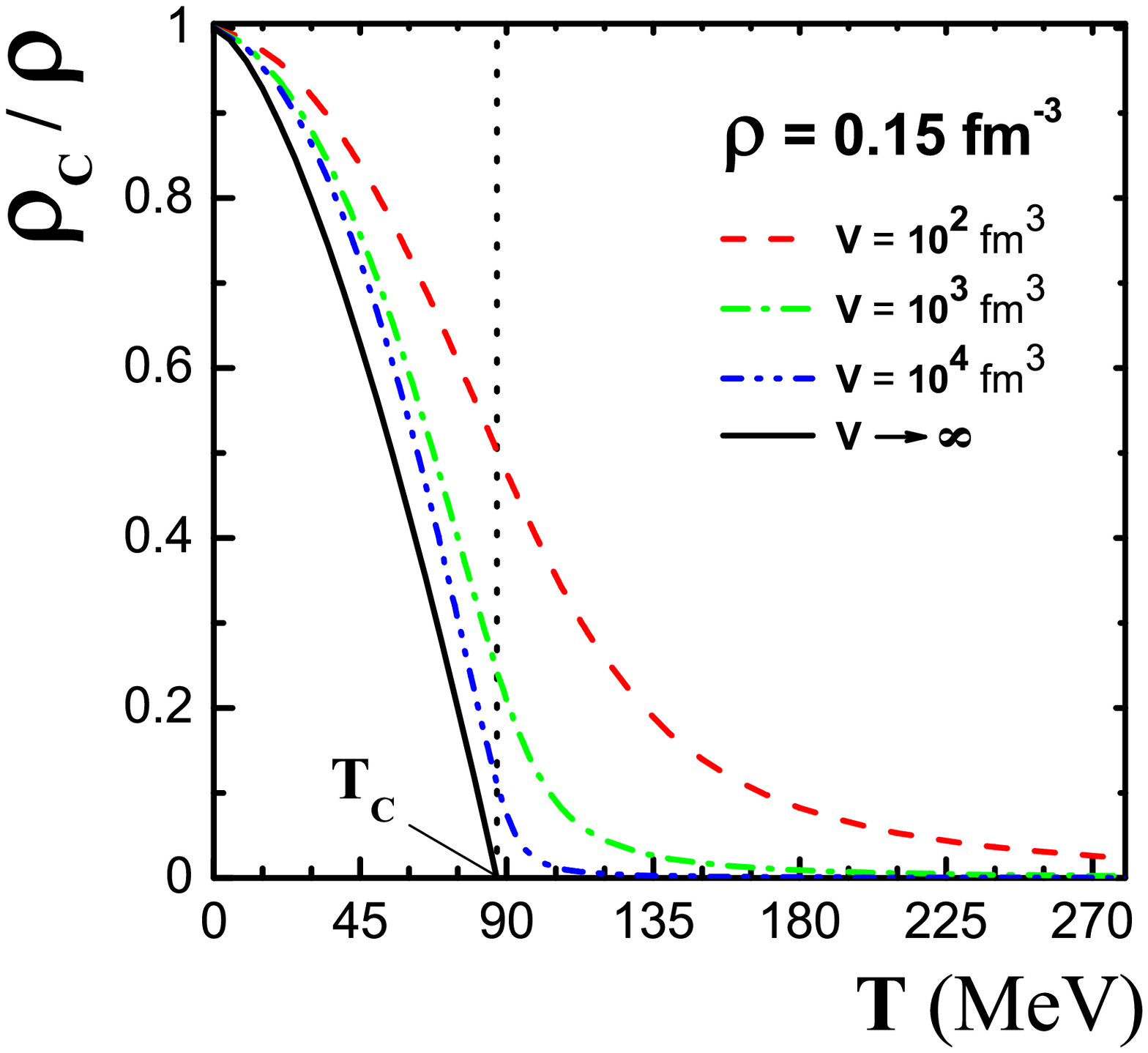,width=0.45\textwidth}
 \epsfig{file=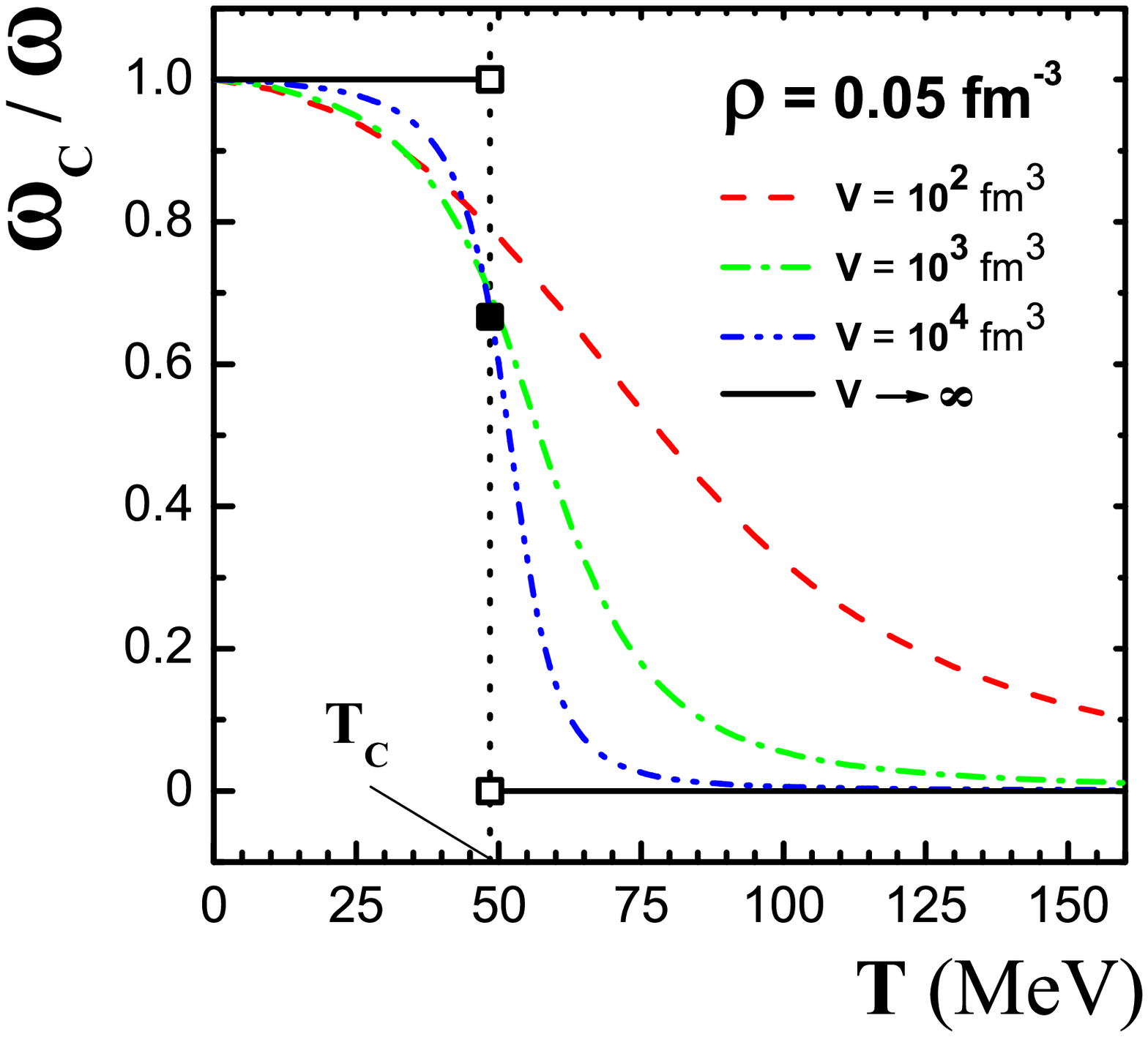,width=0.45\textwidth}\quad
 \epsfig{file=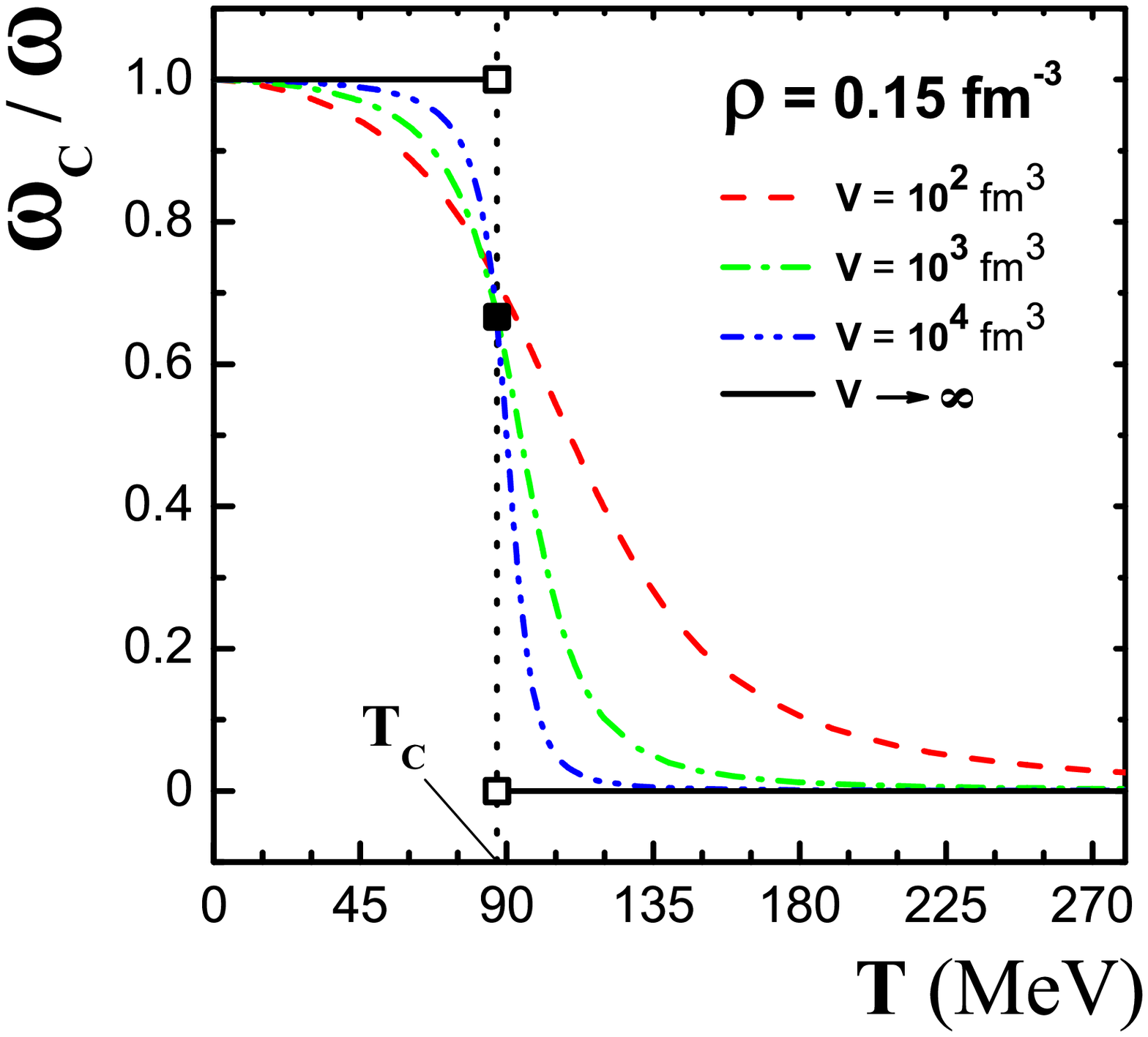,width=0.45\textwidth}
% \\
 \caption{The upper panel shows the ratio of condensate particle number density
 to the total particle number density, $\rho_C/\rho$,
 as functions of $T$ for $V=10^2$, 10$^4$, $10^4$~fm$^3$, and in the TL
 $V\rightarrow\infty$. The lower panel shows the ratio of particle number
 fluctuations in condensate to the total particle number fluctuations, $\omega_C/\omega$,
 as functions of $T$ for the same volumes. The vertical dotted line indicates
 the BEC temperature $T_C$. The left panel corresponds  to $\rho=0.05$~fm$^{-3}$,
 the right one to $\rho=0.15$~fm$^{-3}$~. }\label{fig-C}
\end{figure}

The substitution of $\delta$ in Eq.~(\ref{Wpi0}) from
(\ref{deltaT2}), gives for $T=T_C$ and $V\rightarrow\infty$,
\eq{\label{omegaT3}
\omega~\cong~3~a^{4/3}~\rho^{-1}~V^{1/3}~+~\frac{3}{2}~
a^{4/3}~\rho^{-1}~V^{1/3}~\equiv~\omega_C~+~\omega^*~.
}
The Fig.~\ref{fig-C} demonstrates the ratios $\rho_C/\rho$ and
$\omega_C/\omega$ as the functions of $T$ for $V=10^2$, $10^3$,
$10^4$~fm$^3$, and at $V\rightarrow\infty$. In the TL
$V\rightarrow\infty$, one finds $\rho_C \rightarrow 0$ at $T\ge
T_C$. The value of $\rho_C$ starts to increase from zero at
$T=T_C$ to $\rho$ at $T\rightarrow 0$. Thus, $\rho_C$ remains a
continuous function of $T$ in the TL. In contrast to this, both
$\omega_C$ and $\omega^*$ have discontinuities at $T=T_C$. They
both go to infinity in the TL $V\rightarrow\infty$. The
$\omega_C/\omega$ ratio equals to zero at $T>T_C$, `jumps' from 0
to 2/3 at $T=T_C$, and further continuously approaches to 1 at
$T\rightarrow 0$. At $T=T_C$ the contribution of $p=0$ level to
particle density, $\rho_C$, is negligible at $V\rightarrow\infty$,
but the scaled variance $\omega_C$ from this level equals 2/3 of
the total scaled variance $\omega$ and diverges as $V^{1/3}$. We
conclude this section by stressing that the particle number
fluctuations expressed by the scaled variance $\omega$ looks as a
very promising quantity to search for the BEC in the pion gas.
\section{BEC Fluctuation Signals in High Multiplicity Events}\label{s-BEC-eps}
In the GCE, the scaled variances for different charge pion states,
$j=+,-,0$, are equal to each other and equal to the scaled
variance $\omega$ for total number of pions,
\eq{
\omega^j~
% = \omega^- = \omega^0
 =~ 1 + \frac{\sum_{{\bf p},j}\langle
n_{{\bf p},j}\rangle^2}{\sum_{{\bf p},j}\langle n_{{\bf
p},j}\rangle}~=~\omega~.
\label{omega}
}
There is a qualitative difference in the properties of the mean
multiplicity and the scaled variance of multiplicity distribution
in statistical models. In the case of the mean multiplicity
results obtained with the GCE, canonical ensemble, and
micro-canonical ensemble (MCE) approach  each other in the TL. One
refers here to the thermodynamical equivalence of the statistical
ensembles. It was recently found \cite{CE,MCE,CLT,G-B} that
corresponding results for the scaled variance are different in
different ensembles, and thus the scaled variance is sensitive to
conservation laws obeyed by a statistical system. The differences
are preserved  in the thermodynamic limit. Therefore, the pion
number densities are the same in different statistical ensembles,
but this is not the case for the scaled variances of pion
fluctuations. The pion number fluctuations in the system with
fixed electric charge, $Q=0$, total pion number, $N$, and total
energy, $E$, should be treated in the MCE. The volume $V$ is one
more MCE parameter.

The MCE microscopic correlators equal to (see also
Refs.~\cite{bec2,MCE}):
 \eq{ \label{mcorr}
 \langle \Delta  n_{{\bf p},j} \Delta n_{{\bf k},i}
\rangle_{mce}
 ~ &= ~ \upsilon_{{\bf p},j}^2~\delta_{{\bf p}{\bf k}}\delta_{ji}
 \nonumber \\
 & -~  \upsilon_{{\bf p},j}^2\,\upsilon_{{\bf k},i}^2
 \left[ \frac{q_jq_i}{\Delta (q^2)}
 ~ + ~  \frac{\Delta (\epsilon^2)
 ~  + ~ \epsilon_{\bf{p}}\epsilon_{\bf{k}}\; \Delta (\pi^2)
   -  (\epsilon_{\bf{p}} + \epsilon_{\bf{k}})\Delta (\pi\epsilon)}
 {\Delta (\pi^2)\Delta (\epsilon^2)  -  (\Delta (\pi\epsilon))^2}
 \right]\, ,
  }
where $q_+=1,~q_-=-1,~q_0=0$~,~~
%$\upsilon_{{\bf p},j}^2=
%(1+\langle n_{{\bf p},j}\rangle)\langle n_{{\bf p},j}\rangle$~;~~
%
$ \Delta (q^2) = \sum_{{\bf p},j}q_j^2 \upsilon_{{\bf p},j}^2~,~~
 \Delta (\pi^2) = \sum_{{\bf p},j} \upsilon_{{\bf p},j}^2~,~~
\Delta (\epsilon^2)
 = \sum_{{\bf p},j} \epsilon_{\bf{p}}^2 \upsilon_{{\bf p},j}^2~,~~
 \Delta (\pi \epsilon) =
   \sum_{{\bf p},j} \epsilon_{\bf{p}} \upsilon_{{\bf p},j}^2~$.
Note that the first term in the r.h.s. of Eq.~(\ref{mcorr})
corresponds to the GCE (\ref{np-corr}). From Eq.~(\ref{mcorr}) one
notices that the MCE fluctuations of each mode ${\bf p}$ are
reduced, and the (anti)correlations between different modes ${\bf
p}\ne {\bf k}$ and between different charge states appear. This
results in a suppression  of scaled variance $\omega_{mce}$ in a
comparison with the corresponding one $\omega$ in the GCE. Note
that the MCE microscopic correlators (\ref{mcorr}), although being
different from that in the GCE, are expressed with the quantities
calculated in the GCE.
%The MCE scaled variances depend on two GCE
%parameters: $T$ and $\mu$.
%
The straightforward calculations lead to the following MCE scaled
variance for $\pi^0$-mesons \cite{bec2}:
\eq{
 \omega^0_{mce} ~=~
 \frac{\sum_{\bf{p},\bf{k}}\langle
 \Delta n_{{\bf p},0}~\Delta n_{{\bf k},0}
 \rangle_{mce}}{\sum_{\bf{p}}\langle n_{{\bf p},0}\rangle}~\cong~
 \frac{2}{3}~\omega~.
 \label{omega-mce}
 }
Due to conditions, $N_+\equiv N_-$ and $N_++N_-+N_0\equiv N$, it
follows, $\omega_{mce}^{\pm}=\omega_{mce}^0/4=\omega/6$ and
$\omega_{mce}^{ch}=\omega_{mce}^0/2=\omega/3$, where $N_{ch}\equiv
N_++N_-$.
%The behavior of $\omega^0_{mce}$ is
%shown in Fig.~\ref{fig-w}.

The pion number fluctuations can be studied in high energy
particle and/or nuclei collisions. To search for the BEC
fluctuation signals one needs the event-by-event identifications
of both charge and neutral pions. Unfortunately, in most
event-by-event studies, only charge pions are detected. In this
case the global conservation laws would lead to the strong
suppression of the particle number fluctuations, see also
Ref.~\cite{bec2}, and no anomalous BEC fluctuations would be seen.

\begin{figure}[h!]
 \epsfig{file=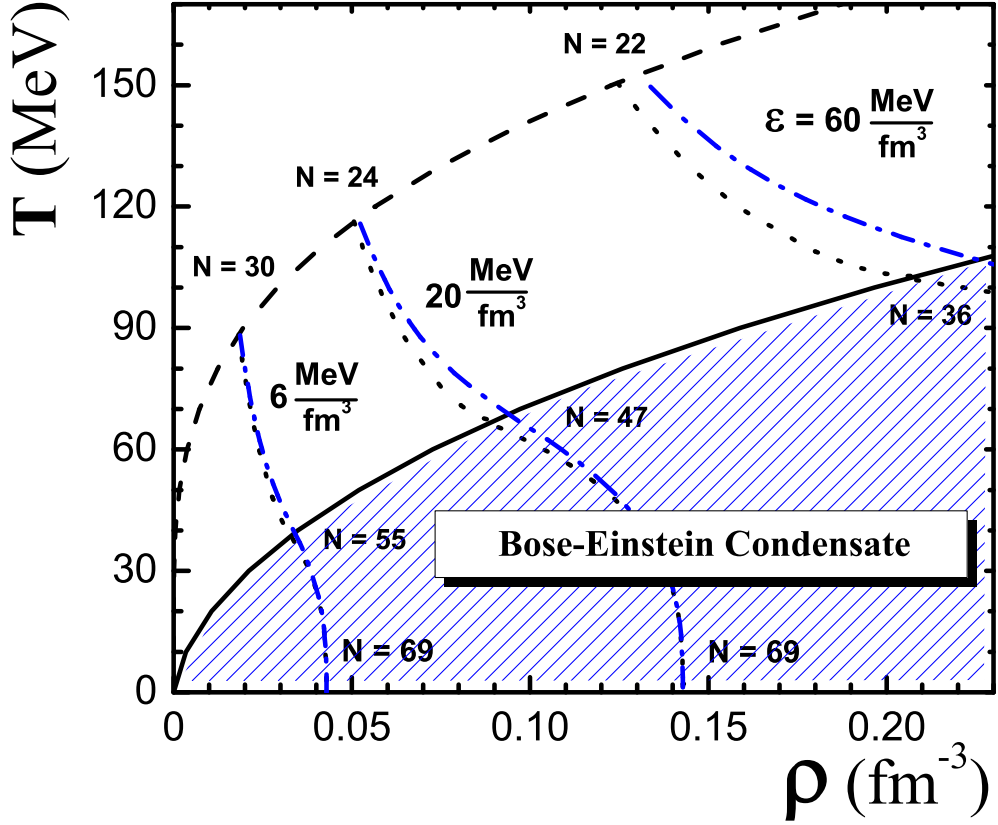,width=0.8\textwidth}
 \caption{The phase diagram of the ideal pion gas with zero net electric charge.
The dashed line corresponds to $\rho=\rho^*(T,\mu=0)$ and the
solid line to the BEC $T=T_C$ (\ref{T_BC}), both calculated in the
TL $V\rightarrow\infty$. The dashed-dotted lines present the
trajectories in the $\rho-T$ plane with fixed energy densities,
$\varepsilon =6,~20,~60$~MeV/fm$^3$, calculated for the finite
pion system with total energy $E=9.7$~GeV according to
Eq.~(\ref{epsV}). The dotted lines show the same trajectories
calculated in the TL $V\rightarrow\infty$. The total numbers of
pions $N$ marked along the dashed-dotted lines  correspond to 3
points: $\mu=0,~T=T_C$, and $T=0$ for $E=9.7$~GeV.
    \label{fig-PD1}}
\end{figure}

As an example we consider the high $\pi$-multiplicity events in
$p+p$ collisions at  the beam energy of 70~GeV (see
Ref.~\cite{Dubna}). In  the reaction $p+p\rightarrow p+p+N$ with
small final proton momenta in the c.m.s.,  the total c.m. energy
of created pions is $E\cong \sqrt{s}-2m_{p}\cong 9.7$~GeV. The
estimates~\cite{nikitin} reveal a possibility to accumulate the
samples of events with fixed $N=30\div 50$ and have the full pion
identification. Note that for this reaction the kinematic limit is
$N^{max}=E/m_{\pi}\cong 69 $. To define the MCE pion system one
needs to assume the value of $V$, in addition to given fixed
values of $Q=0$, $E\cong9.7$~GeV, and $N$. The $T$ and $\mu$
parameters of the GCE can be then estimated from the following
equations,
\eq{\label{EVN}
E~=~V~\varepsilon(T,\mu;V)~,~~~~N~=~V~\rho(T,\mu;V)~.
}
In calculating the $\varepsilon$ and $\rho$ in Eq.~(\ref{EVN}) we
take into account the finite volume effects according to
Eqs.~(\ref{rhoV}-\ref{epsV}) as it is discussed in Sec. II.
Several `trajectories' with fixed energy density  are shown in
Fig.~\ref{fig-PD1} starting from the line $\mu=0$  in the pion gas
in the $\rho-T$ phase diagram. The MCE scaled variance of $\pi^0$
number fluctuations, $\omega_{mce}^{0}$, increases with increasing
of $N$. The maximal value it reaches at $T\rightarrow 0$,
 \eq{
 \omega^{0~max}_{mce}
 \;\cong\; \frac{2}{3}~\left(1~+~\langle N_0\rangle^{max}\right)~
 =~\frac{2}{3}~\left(1~+~\frac{~N^{max}}{3}\right)~\cong~16 ~.
 }
\begin{figure}[h!]
 \epsfig{file=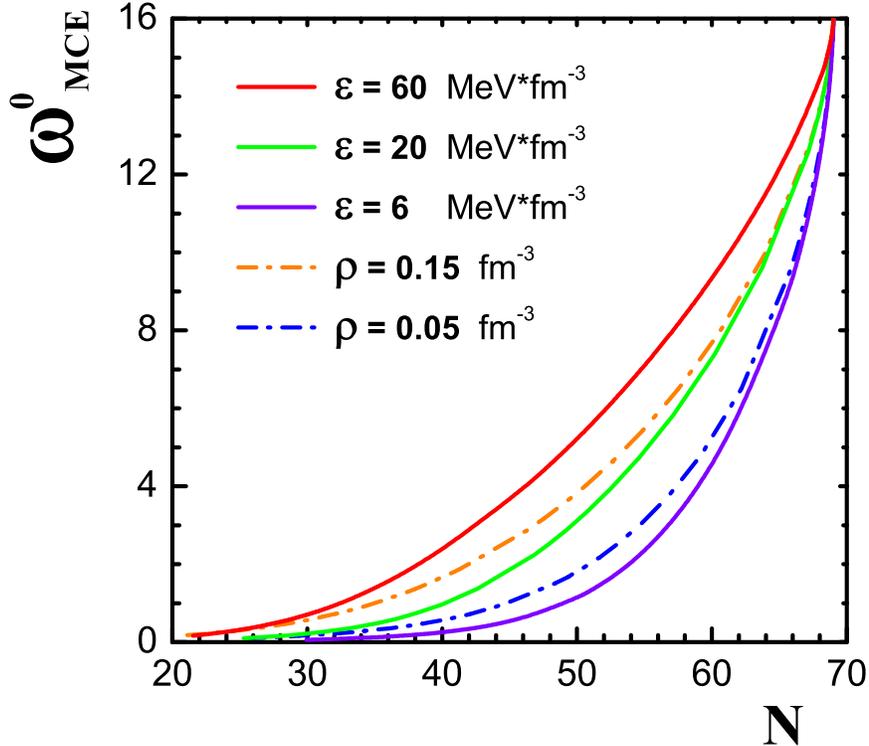,width=0.5\textheight}
 \caption{
The scaled variance of neutral pions in the MCE is presented as
the function of the total number of pions $N$. Three solid lines
correspond to different energy densities, $\varepsilon~
=~6,~20,~60$~MeV/fm$^3$ (from bottom to top), calculated according
to Eq.~(\ref{epsV}). Two dashed-dotted lines correspond to
different particle number densities, $\rho~
=~0.05,~0.15$~fm$^{-3}$ (from bottom to top), calculated according
to Eq.~(\ref{rhoV}). The scaled variance $\omega^0_{mce}$ is given
by Eq.~(\ref{omega-mce}), with $\omega$ (\ref{W_def}) and $\langle
\Delta N^2 \rangle$ (\ref{DN2}). The total energy of the pion
system is assumed to be fixed, $E=9.7$~GeV.
 \label{fig-omega0}}
\end{figure}

In Fig.~\ref{fig-omega0}, $ \omega^0_{mce}$ is shown as the
function of $N$. Different possibilities of fixed energy densities
and fixed particle number densities are considered. One way or
another, an increase of $N$ leads to a strong increase of the
fluctuations of $N_0$ and $N_{ch}$ numbers due to the BEC effects.

The large fluctuations of $N_0/N_{ch}=f$ ratio were also suggested
(see, e.g., Ref.~\cite{DCC}) as a possible signal for the
disoriented chiral condensate (DCC). The DCC leads to the
distribution of $f$ in the form, $dW(f)/df=1/(2\sqrt{f})$. The
thermal Bose gas corresponds to the $f$-distribution centered at
$f=1/2$. Therefore, $f$-distributions from BEC and DCC are very
different, and this gives a possibility to distinguish between
these two phenomena.

\section{Summary}
The idea for searching the pion BEC as an anomalous increase of
the pion number fluctuations was suggested in our previous paper
\cite{bec2}.  The fluctuation signals of the BEC have been
discussed in Ref.~\cite{bec2} in the thermodynamic limit. At
$V\rightarrow\infty$, it follows, $\omega=\infty$ at $T\le T_C$.
This is evidently not the case for the finite systems. At finite
$V$ the scaled variance $\omega$ of the pion number fluctuation is
finite for all possible combinations of the statistical system
parameters. The $\omega$ demonstrates different dependence on the
system volume $V$ in different parts of the $\rho-T$ phase
diagram. In the TL $V\rightarrow\infty$, it follows that $\omega$
converges to a finite value at $T>T_C$. It increases as
$\omega\propto V^{1/3}$ at the BEC line $T=T_C$, and it is
proportional to the system volume, $\omega\propto V$, at $T<T_C$.
The statistical model description gives no answer on the value of
$V$ for given $E$ and $N$. The system volume remains a free model
parameter. Thus, the statistical model does not suggest an exact
quantitative predictions for the $N$-dependence of
$\omega^0_{mce}$ and $\omega^{\pm}_{mce}$ in the sample of high
energy collision events. However, the qualitative prediction looks
rather clear: with increasing of $N$ the pion system approaches
the conditions of the BEC. One observes an anomalous increase of
the scaled variances of neutral and charged pion number
fluctuations. The size of this increase is restricted by the
finite size of the pion system. In turn, a size of the created
pion system (maximal possible values of $N$ and $V$) should
increase with the collision energy.

\begin{acknowledgments}
We would like to thank  A.I.~Bugrij, M.~Ga\'zdzicki, W.~Greiner,
V.P.~Gusynin, M.~Hauer, B.I.~Lev, St.~Mr\'owczy\'nski,
M.~Stephanov, and E.~Shuryak for discussions. We are also grateful
to E.S.~Kokoulina and V.A.~Nikitin for the information concerning
to their experimental project \cite{Dubna}. The work was supported
in part by the Program of Fundamental Researches of the Department
of Physics and Astronomy of NAS Ukraine. V.V. Begun would like
also to thank for the support of The International Association for
the Promotion of Cooperation with Scientists from the New
Independent states of the Former Soviet Union (INTAS), Ref. Nr.
06-1000014-6454.
\end{acknowledgments}
%
%\begin{figure}[h!]
% \epsfig{file=N(T)_e20.eps,height=0.3\textheight}\,
% \epsfig{file=W0.eps,height=0.3\textheight}
%
% \caption{\label{fig-N,W}}

%\end{figure}
%

%The another very interesting result is that the inclusion of
%$p_0=0$ level makes scaled variances finite, in contrast to the
%Ref.~\cite{bec2}.

\end{document}